\documentclass[12pt]{article}
\newcommand{\bq}{\begin{equation}}
\newcommand{\eq}{\end{equation}}
\newcommand{\ds}{\displaystyle}
\newcommand{\lng}{\ln g}
\newcommand{\td}{\tilde{d}}
\begin{document}
   \begin{center} {\Large    Schwarzschild  Black Hole
  Quantum Statistics \\ from $Z(2)$ Orientation Degrees of Freedom
  \\ and its Relations to Ising Droplet Nucleation\\[1.0cm] }
{\large \ H.A. Kastrup\footnote{E-Mail:
kastrup@physik.rwth-aachen.de}} \\ Institute for Theoretical
Physics, RWTH Aachen \\ D-52056 Aachen,
 Germany
\end{center} \vspace*{0.6cm}
  {\large  Abstract} \\[0.2cm]

   Generalizing previous quantum gravity results for
    Schwarzschild black holes from 4
    to $D\geq4$ space-time dimensions yields an energy
   spectrum $E_n = \alpha \,n^{ (D-3)/(D-2)} \,E_{P,D}~,~n=1,2, \ldots~,
    \alpha =
   O(1)~, $ where $E_{P,D}$ is the Planck energy in that  space-time.
   This energy spectrum means that the quantized
    area $A_{D-2}(n)$ of the $D-2$ dimensional horizon has universally
    the form  $A_{D-2} =n\,a_{D-2}$, where $a_{D-2}$ is essentially the
    $(D-2)$th power of the $D$-dimensional Planck length.
    Assuming that
 the basic area quantum has a $Z(2)$-degeneracy according to its two
 possible orientation degrees of freedom implies
   a degeneracy $d_n =2^n$ for the $n$-th level.
The energy spectrum with such a degeneracy leads to a {\em
quantum}
 canonical partition
   function which is the same as the {\em classical} grand canonical
   potential of a primitive Ising droplet nucleation
   model for 1st-order phase transitions in $D-2$ spatial dimensions.
   The analogy to this model suggests that  $E_n$ represents
   the surface energy of a "bubble" of $n$ horizon area quanta.
    Exploiting the well-known
   properties of the so-called critical droplets of that model immediately
   leads to the Hawking temperature and the Beken\-stein-Hawking entropy of
   Schwarzschild black holes. The values of temperature and
    entropy  appear closely    related to the imaginary
    part of the partition function which describes metastable  states.
  \newpage
   \section{Introduction}
    In  previous papers \cite{ka1,ka2,ka3} I discussed the
  quantum statistics of the energy spectrum
   \bq E_n = \alpha \, \sqrt{n}\, E_P\;,~ n=1,2, \ldots,~~ E_P=
    c^2 \sqrt{c\, \hbar/G}~,~
   \alpha=O(1)~~, \eq with
    degeneracies \bq d_n = g^n~,~g>1~~. \eq Many authors have proposed
   that  Schwarzschild black holes in 4-dimensional
     space\-times have such a spectrum (see the list in Ref.\ [1]).

      The (quantum) canonical
    partition function of the above spectrum not only leads
     to the Haw\-king temperature
    and the associated Bekenstein-Haw\-king entropy, but, very amazingly,
     it is
   formally the same as the {\it classical} grand canonical potential
     of the primitive droplet nucleation model in the context of first-order
     phase transitions in {\it two} space dimensions. Thus the so-called
     "holographic principle" \cite{tho1,su,tho2}, namely that the essential
 physics of
     black holes should be associated with the 2-dimensional horizon,
     is very evident here and comes out as a result!

     Furthermore, as the canonical partition function of the spectrum (1)
     becomes complex for the degeneracies (2), because $g>1$, one leaves
     the well-established frame\-work of equilibrium thermodynamics (KMS
     states)
     and
     moves on the perhaps more slippery ground of metastable states and
     nonequilibrium thermodynamics \cite{le1}.

  The paper is organized as follows: I first briefly summarize postulates
  and arguments from previous papers \cite{ka1,ka2,ka3,ka4}
     generalizing the  spectrum (1) to $ D \geq 4$ space-time
     dimensions \cite{ka3}.

      Furthermore, I shall argue in chapter 3
      that the much debated
     degeneracy (2) may be attributed to the $Z(2)$ degrees of freedom
      associated with
     the two possible orientations  of
      the basic area quantum , i.e.\ $g=2\,$! This is the
     main new idea of the present paper, namely that the huge entropy of
     black holes is related to the $Z(2)$-valued geometrical quantity
     "orientation".

     Moreover,  it will be shown that for all $ D \geq 4$
     the resulting $quantum$
     canonical partition function is the same as the $classical$ grand canonical
     potential of  the (nonrelativistic)
      primitive droplet nucleation model in $D-2$ spatial
     dimensions, too. The essential features of the droplet nucleation model
     needed for our purpose are discussed
     in chapter 4. From a comparison of the two models one sees that the
     energy $E_n$ in (1) and its $D$-dimensional generalizations are to be
     interpreted as the surface energies of  "bubbles" of $n$ area quanta.

      Using known results from the droplet nucleation model -
     especially the notion of a critical droplet -  the
     Hawking temperature and the Bekenstein-Hawking entropy are derived
     in chapter 5,
     up to a normalization factor which is  discussed separately. In chapter 6
     it is pointed out
  how an effective Hamiltonian of the Born-Infeld type can be
     used to describe  the spectrum plus
     the degeneracy factor. Its
     classical mechanical counterpart has some amusing properties, too.
     Here it is interesting that the imaginary part of the corresponding
     classical partition function is the same as that of the
     quantum mechanically one. The real parts are different, however.
     Chapters 4-6 rely strongly on Ref. \cite{ka3}.

     \section{The quantum Schwarzschild BH spectrum  in $D$-dimensional
     space-time} The spectrum (1), already discussed in Refs.\
      \cite{ka1,ka2,ka3},
      may be argued for heuristically as follows \cite{ka4}: A canonical
     Dirac-type treatment of spherically symmetric pure Einstein gravity
     leads to a reduced 2-dimensional phase space having only the ADM mass
     $M$ and a canonically conjugate time functional $T$ as (observable)
     pair of variables \cite{ka5}.

      In this simple model an observer at spatial (flat) infinity
     will only have the mass $M$ and his own proper time $\tau$ available in
     order to describe the system. His very simple Schr\"odinger equation for
     it is
      \bq i\hbar \partial_{\tau} \phi(\tau) =Mc^2 \phi(\tau)~, \eq
which has the plane wave solutions \bq \phi(M,\tau)= \chi(M)e^{\ds
 -\frac{i}{\hbar}Mc^2\tau}~, \eq where $M > 0 $ is assumed. If the system
 with  mass $M$ stays there forever, then $M$ is to be considered as a
 continuous quantity. However -- just like in the case of plane waves
  representing a system confined to a
 1-dimensional spatial interval with finite extension $L$ -- , if the above system,
  represented by the plane wave (4),  has only a finite
 duration $\Delta$ then this property may be (crudely) implemented by imposing
 periodic boundary conditions on the plane wave (4), implying the
 relation \cite{ka4}
  \bq c^2M\Delta = 2\pi \hbar~ n~,~ n=1,2,
 \ldots~.
   \eq  As to the boundary condition (5) it is worthwhile to stress
    the following point: The assumption that
    the wave  function (4) has the period $\Delta$ does not mean that the
    asymptotic time $\tau$ is periodic. It merely means that the system
    is in a (quasi-) stationary state (4) during the time interval $\Delta$.
    This is completely analogous to a system of free particles in
    a finite spatial interval of length $L$ where they are described
     by a plane wave with periodic spatial boundary conditions.
     Such a property of the wave function does not mean that space
     itself is confined to an interval or periodic. \\
    The question now is how to choose $\Delta$! As the only intrinsic
    quantity
   available at spatial infinity to characterize a
    time interval is $M$ itself, or
   a function of it, namely the Schwarzschild radius $R_S(M)$, we
   assume \cite{ka4}
   \bq \Delta=\gamma R_S(M)/c~, \eq
   where $\gamma$ is a dimensionless number of order 1. The choice (6) means
   that we deal with the (quasistationary) "formation" period of the black
    hole \cite{ka4}, not
   with the Hawking evaporation period which, according to the law of
   Stefan-Boltzmann has the different time scale $\propto R_S^{D-1}$.
   Details  will be discussed at the end of chapter 5.

   Inserting the ansatz (6) into relation
   (5) gives the mass quantization condition
    \bq \gamma cM_n\, R_S(M_n) = 2\pi \hbar~ n~,~ n=1,2,
 \ldots~.
   \eq  For $D=4$ we have $R_S=2MG/c^2$ and the spectrum (1) results.
   However, one can assume the relation (7) to be valid in any dimension
    $D\geq 4$, because
   the Schr\"odinger Eq.\ (3) has to hold in any such dimension \cite{bo1}!

The relation (7) is also an appropriate
   generalization of a Bohr-Sommer\-feld type quantization of the 2-dimensional
   horizon as suggested very early by Bekenstein \cite{be1},
     Mukhanov \cite{mu1}
   (see also  Bekenstein's recent review \cite{be2}) and -- in the context of
   string theory -- by Kogan \cite{ko}:

   If we interpret $cM$ as canonical momentum and $R_S$ as canonical
   coordinate, then the relation (7) is,  qualitatively,  nothing else but the
   old-fashioned quantum counting of phase space cells.

   Those readers who are - at present - skeptical about the arguments
   leading to the Eqs.\ (5) and (6) may take them as mere
   assumptions which - together with the degeneracies discussed
   below -
    lead to interesting physical consequences including those
    generally expected from black holes. The close correspondence
    to the droplet nucleation model to be discussed below will
    shed additional light on the physical significance of those
    assumptions.

   In the following it is convenient to use these notations: We
   put $D=1+d=2+ \tilde{d}$: $d$ gives the number of space dimensions and
   $\tilde{d}$ the spatial dimensions of the black hole horizon.

   In $D$ space-time dimensions the Schwarzschild radius is given by
    \cite{pe1} \bq
   R_S(M) = \left(\frac{16 \pi G_D M}{c^2\,
    \omega_{\td}\;\td}\right)^{ 1/(\td-1)}~, \eq where $G_D$ is the
   gravitational
    constant in $D$-dimensional space-time and \[ \omega_{\td} = 2 \pi^{(\td
    +1)/2}/ \Gamma( (\td +1)/2)\] is the volume of $S^{\td}$. Inserting
    this $R_S$ into the relation (7) gives the energy spectrum
    \begin{eqnarray}E_n = M_n c^2 & =& \alpha_D\, n^{ 1-\eta}\, E_{P,D}~,~~
     \eta=1/\td~, \\
   \alpha_D &=&  \left(\frac{(2\pi)^{\td -2}\;  \omega_{\td}\;\td}{8\,
    \gamma^{\td
    -1}}\right)^{\ds \eta}~\approx O(1)~,
     \nonumber \end{eqnarray} where \bq  E_{P,D} =
    \left(\frac{\hbar^{D-3} c^{D+1}}{G_D}\right)^{1/(D-2)}~,~~~
    l_{P,D}=\left(\frac{\hbar G_D}{c^3}\right)^{1/(D-2)} \eq are the
    corresponding Planck energy  and Planck length in $D$ space-time
    dimensions, respectively.

     Combining
    the Eqs.\ (7) and (8) yields the following important expression for the
     $\td$-dimensional  quantized
    area of the horizon \bq A_{\td}(n)= (R_S(M_n))^{\td}\,
     \omega_{\td} = n\, a_{D-2}\,,~~ a_{D-2}=\frac{32\,
     \pi^2}{\gamma \td} \, l_{P,D}^{\td}~~,~~ n=1,2, \ldots~. \eq
     Thus, the horizon is built up additively and equidistantly
     from Planck-sized area elements $a_{D-2}$ universally for
     all space-time dimensions $D\geq 4$!

     Notice that
     \[A_{\td}(n_1+n_2)=A_{\td}(n_1)+A_{\td}(n_2)~,\]but
     (Minkowski's inequality)
     \[E_{n_1 +n_2}< E_{n_1}+E_{n_2} \mbox{ for } n_1,n_2 \geq
     1~. \]

     Therefore it is energetically advantageous to form $one$ large
     area instead of several smaller ones, the energy difference
     being "radiated" away! Thus, the following picture emerges (see also
     chapter 5): If two spherically symmetric  horizon quantum
     "bubbles",
     characterized by the numbers $n_1$ and $n_2$, merge, then the area of
     the resulting quantum bubble is the sum of the area of the original
      surfaces whereas the resulting surface energy is lower than
      the sum of the original ones.
      This is a kind of quantum version of Hawking's  classical
      area theorem \cite{Ha0}.

     The spectra (11) and (9), respectively, can be
    substantiated considerably by an appropriate group theoretical
    quantization \cite{is} of the classical system associated with the
    symplectic form $d \tau \wedge dM $ (which is symplectically
    equivalent to
    $d\varphi \wedge dp,~\varphi \in (-\pi, +\pi],~p>0$)  in terms of the group
     $SO^{\uparrow}(1,2)$
    and the positive discrete series of its unitary representations \cite{bo1}.

    The main result of this group theoretical quantization is that
    \bq A_{\td}(k;\tilde{n}) \propto k+\tilde{n}~, \tilde{n}
    =0,1,2,\ldots~,\eq where $k=1,2,\ldots$
    characterizes the irreducible unitary representation
    of $SO^{\uparrow}(1,2)$. Physically $k$  determines the energy value of the
    ground state for which $\tilde{n}=0$.
    The unitary representation corresponding to the spectrum (11) belongs
    to $k=1$ and can be
    realised by complex-valued functions $f(z), |z=x+iy|=\sqrt{x^2+y^2}<1$, with
    the scalar product \bq (f_1,f_2)=\frac{1}{\pi}\int_{|z|<1}dxdy
    \bar{f}_1(z)f_2(z) \eq and the orthonormal basis \bq \phi_{\tilde{n}}
    (z)=\sqrt{\tilde{n}+1}\, z^{\tilde{n}}~,~\tilde{n}=0,1,2, \ldots
    ~~.\eq
    The operator with the eigenvalues
    $n=\tilde{n}+1$ and the eigenfunctions $\phi_{\tilde{n}}$ is $zd/dz+1$.
      For  more details see Ref.\ \cite{bo1}. Notice
    that the eigenfunctions $\phi_{\tilde{n}}, ~\tilde{n}>1,$ are -- up to the
    normalization -- just powers of
    $\phi_1$! This is re\-mi\-niscent of a Fock space structure: the
    $\tilde{n}$-area quantum wave function is essentially the $\tilde{n}$-th
    power of the $1$-area quantum wave function!

    It is an interesting and encouraging result \cite{bo2} that for $D=4$ the
    (horizon) area operator in the spherically symmetric sector of loop quantum
    gravity has eigenvalues which for large $n$ are proportional
    to $n$, too.
    \section{Degeneracy of the spectrum due to the \\
    $Z(2)$-freedom of orientation}
    I would like to stress
    that a relationship like (2) for the degeneracies is very
    important for the thermodynamics
    involved \cite{ka1,ka2,ka3}! Like the spectrum (1) the
      ansatz (2) for the degeneracies has a
    longer history, too:

     It is already implicitly contained in Bekenstein's
    early work (see his review \cite{be3}).  Afterwards
    Zurek and Thorne \cite{zu}
    interpreted the degeneracies as the number of
  quantum mechanically distinct ways the black hole could  be made
  by infalling quanta. If their total number is $n$ then the associated
  combinatorial sum
  $$ \sum_{k=0}^n {n \choose k} = 2^n $$ yields the degeneracy (2) with
  $g=2$. Similarly, Mukhanov suggested \cite{mu1} to take as $d_n$  the
  number of possible ways to build up the $n$-th level if the ground state
  is nondegenerate. This gives $d_n =2^{n-1}$. 't Hooft \cite{tho3,tho2}
  pointed out that one gets the right relation between the area of
   the horizon and the entropy if one divides the horizon into Planck-sized
   cells and attributes two degrees of freedom (i.e.\ a $Z(2)$-symmetry)
 to each cell.
  Sorkin and others \cite{so1,so2} \cite{sc} have argued similarly.
   See also the recent review by Bekenstein \cite{be2}.

   The big question then is, {\em where does
  such a degeneracy in terms of a microscopic $Z(2)$-symmetry come from
   geometrically?}
  If it is already present for a pure Einstein quantum gravity
   system -- uncoupled to matter
  -- it has to have a geometrical origin!

   Actually there is such a geometrical
  degree of freedom, namely {\em orientation}: Most of the
   physically interesting
  manifolds are orientable (see the Appendix), having exactly two equivalence
  classes $[\omega^{n}_{\pm}=f_{\pm}(x) dx^1\wedge \ldots
  \wedge dx^n,~ f_+(x)>0,
  f_-(x) < 0]\equiv \sigma_M= +1,-1$ of everywhere nonvanishing
   $n$-forms $ \in \Lambda^n(M)$
   of a $n$-dimensional manifold $M$. The convention is that the $R^n$ with a
   natural ordering of its basis has $\sigma_{R^n}=\sigma[e^1\wedge \ldots
   \wedge e^n]=+1$ and that this
   orientation is induced on $M$ by the orientation-preserving atlas of
   coordinate
   systems.

    Orientation is a global property of a classical
   manifold, and
   -- similar to space reflection parity in atomic and particle physics
   --
   it appears suggestive to attribute this degree of freedom to the smallest
   area quanta of the corresponding quantum gravity system, too. Here the
   classical system is the horizon bifurcation sphere $\Sigma^{D-2}$ which is
   orientable (see Appendix). \\ So, if we associate with each area quantum
   $a_{D-2}$ in Eq.\ (11) a $Z(2)$-degree of free\-dom $\sigma =\pm 1$ then a
   state with $n$ such quanta has a degeneracy $2^n$. Because of the
   properties of "orientation"
    this degeneracy is universally the same for all dimensions $D$.

   As the two orientational
   degrees of freedom of the elementary area quanta
   appear to be energetically undistinguishable -- at least
   in a first approximation -- they lead to the immense degeneracy (2)
   implying
   the Bekenstein entropy which is very much larger than that associated
   with the mere two orientation
   degrees of freedom of a classical geometrical sphere.

   The two properties -- the quantum statistical and the classical one --
   may be related as follows:
    Suppose there are -- in a second approximation --
   possible new microscopic (weak) attractive interactions between the
   elementary
   orientation degrees of freedom, perhaps similar
   to  the exchange
   interactions between spins due to Fermi-Dirac statistics, or more likely,
    similar to
   "attractions" due to Bose-Einstein statistics (with condensation).

    Actually, the
   variable $\sigma(i)$ of the $i$-th horizon area quantum is an
    Ising-type variable
   and there is the question whether it shares other properties
    of the Ising
   model \cite{do1}:

   At high (Hawking) temperatures (black holes with small masses)
   the microscopic
   Ising variables form a ("para"-) phase of non-correlated Ising
   "spins".

    But then there may be  something like "spontaneous orientation" at
   very low temperatures! Recall that for the 2-dimensional Ising model
   on a square lattice we have the relation $J/k_BT_c \approx 0.4$ between
   the critical temperature $T_c$ and the coupling $J$ of nearest neighbors.
   As the Hawking temperature $T_H$ of macroscopic black holes
    ($D=4$) with masses of about
    10 solar masses is of the order $10^{-8} K \approx 10^{-12}~ eV$,
   then the energy scale $J$ is extremely small $if$ the corresponding
   Ising $T_c$ is of an
    order similar to that of $T_H$.

     The possible occurrence of a spontaneous
    orientation at very low temperatures  of very large black holes could
    explain why we can speak of a macroscopic classical horizon the associated
    sphere of which has just two possible orientations
    corresponding
    to the classical limit $\hbar \rightarrow 0, T_H \rightarrow 0$.

    In any case it is very important to clarify the problem
    whether the geometrical degree of freedom "orientation" is merely a
    kinematical one or whether it has dynamical properties, too,
    and how those are related to Einstein's gravity.

     What could correspond to the external magnetic
   field in our case?

   Classically
   a suitable representative of the equivalence class $\sigma$ which
   is invariant under orientation-pres\-erving dif\-feo\-mor\-phisms is
    the volume
   element $\sqrt{\gamma^{(\td)}(u)}\;du^1\wedge \ldots \wedge du^{\td}$ of
   $\Sigma^{\td}$, where $\gamma^{(\td)}$ is the determinant of the induced
   metric on $\Sigma^{\td}$. Diffeomor\-phism-invariant $external$ fields
   are the curvature scalar $^{\td}R$, a (cosmological) constant or any
   other  classical diffeomorphism invariant function, which even might be
   related to matter fields, e.g. $F^{jk}F_{jk}(u)$. The total contribution
   of all volume elements is then given by the integral over
   $\Sigma^{\td}$. Thus, if, e.g.\ the external field is a constant and if
   $\Sigma^{\td}$ is a sphere, then the total ''interaction'' is proportional
   to the area of $\Sigma^{\td}$! The situation is  similar to the
   coupling of $p$-branes to $(p+1)$-forms in the context of
   theories of strings and $p$-branes \cite{ka7,po}.

    Quantum theoretically the above volume elements should be
    replaced by  area quanta "coupled" to some classical surface
    and the integral replaced by a corresponding sum. In order to
    work this out one has to know more about the "interactions"
    of the orientation degrees of freedom, similar to the
    interactions of atomic spins (magnetic moments) in para- and
    ferromagnets.

    Taking the Ising orientation degrees of freedom $\sigma(i)=\pm $
    into account
    implies that the wave function (14) is to be replaced by
    \[\phi_{\tilde{ n}}(z)\; \chi(\sigma(0), \sigma(1),\ldots,
    \sigma(\tilde{n}))~,~\tilde{n}=0,1,\ldots, \]
    where $\chi$ is symmetric with respect to the exchange
    of any pair of the $\sigma(i)~,~i = 1,\ldots \tilde{n}$.

    These few remarks indicate that introducing the Ising
   spin type variable "orientation" as a new degree of freedom opens a
   wide range of new theoretical possibilities for analyzing black holes and
   other gravitational systems. Here we shall merely exploit the fact that
   the spectrum (9) corresponds to  surface energies of  droplets of $n$ Ising
   spins in a background of opposite Ising spins (see the next chapter).

    Another point to comment on is the free parameter $\gamma$.
     We shall see below how
   the Gibbons-Hawking geometrical approach \cite{gi1} to the
     partition function may be used to restrict it, but it is
      worth mentioning
     that such a free parameter also occurs in loop quantum gravity
      \cite{as1},
      in the discussion of the Schwarzschild black hole thermodynamics
     of  Matrix theory \cite{pe} and in the AdS/CFT correspondence \cite{wi}.

     In a microcanonical approach to the thermodynamic properties
     of the system we have for the entropy \bq S(n)/k_B = \ln d_n =
     n\, \ln 2 = \frac{\ln 2}{a_{D-2}}~ A_{D-2}(n). \eq If we define, in
     analogy to the classical expression $T=\partial U/\partial
     S,~U:$ internal energy,  for sufficiently large $n$, \bq
     T \approx \frac{E_{n+1}-E_n}{S(n+1)-S(n)} \eq we get \bq T \propto
     n^{-\eta} \eq which corresponds to the classical relationship
      $T \propto 1/R_S (M)$ \cite{pe1}.

       The quantum canonical partition function resulting from the
      degeneracy (2) and the spectrum (9) may
     be written as
     \begin{eqnarray} Z_D(t,x)& =&
   \sum_{n=0}^{\infty}e^{\ds n t -n^{1-\eta}x}~, \\
  & &t= \ln g~,~~ \eta = 1/\td \equiv 1/(D-2)~, \nonumber \\
  & &  x= \beta \,\alpha_D\, E_{P,D}~,~~\beta = \frac{1}{k_B T}~.
   \nonumber \end{eqnarray} It is important for the following
  discussion to keep $g$ arbitrary and put $g=2$ only in some final results.
 One reason is that the series (18) does not converge for $g>1$ and one has to
 make analytical continuations. In the following it is
  mathematically convenient to start
 the sum (18) at $n=0$ instead of $n=1$.

  The function
   $Z_D(t,x)$ obeys the linear PDE
 \bq \partial_t^{\td-1}Z_D = (-1)^{\td} \,\partial_x^{\td}
    Z_D, \eq which can be used to determine $Z_D$ in closed form. This has
    been done \cite{ka6} for $\td=2$ and $3$.
  \section{Primitive droplet nucleation model \\  in $D-2$ space
 dimensions}
  The model assumption that 1st-order phase transitions
   are  initiated by the
 formation (homogeneous nucleation) of expanding droplets
  of the new phase within the old phase is a popular and important
  one (see the reviews \cite{rev} with their references to the original
  literature).

  In its most primitive form the droplets are
  assumed to be spherical and  to
  consist of $n$ constituents (e.g.\ droplets of Ising spins on a lattice or
  liquid droplets of molecules etc.),
  the "excess" energy $\epsilon_n$ of which is given by a  "bulk"
   term proportional to the volume
  $n$ and a  term proportional to the surface $n^{1-\eta}, \eta =1/\td$, where
  $\td \ge 2$ is the spatial dimension of the system:  \bq   \epsilon_n =
   -\hat{\mu}\, n +
  \phi \, n^{1-\eta}~,~~\eta =1/\td~.\eq In the case of negative Ising spin
  droplets, formed in a background of positive spins by turning
  an external magnetic field
  $H$ slowly negative - below the critical temperature -, the coefficient
  $\hat{\mu}$ in Eq.\ (20) takes the form $\hat{\mu}=-2H$ and for liquid
 droplets of $n$
  molecules condensing from a super-saturated vapour one has $\hat{\mu}= \mu
  -\mu_c$, where $\mu$ is the chemical potential and $\mu_c$ its critical
  value at condensation point. $\phi$ is the constant surface tension.\\
   Assuming the average number $\bar{\nu}(n) $ of
  droplets with
  $n$ constituents to be proportional to a Boltzmann factor, \bq
    \bar{\nu}(n)
 \propto e^{\ds -\beta
  \epsilon_n},~ \beta = \frac{1}{k_B T}~, \eq and  that the droplets form
  a non-interacting dilute gas leads to the grand canonical
   potential $\psi_{\td} $
  per spin or per
  volume \begin{eqnarray}
  \psi_{\td} (\beta, t = \beta \hat{\mu}) & =& \ln Z_G  = p \beta =
   \sum_{n=0}^{\infty}e^{\ds  t n-x n^{1-\eta}}~, \\ & & t=\beta
   \hat{\mu},~ x=\beta \phi~;~ p:\mbox{pressure}, \nonumber \\ & &
 d\psi_{\td} = -U d \beta +
    \bar{n}dt~~ . \end{eqnarray}
  (Again: for physical reasons  the sum (22) may not start at $n=0$ but at some
  finite $n_0 >0$. This can easily be taken care of.)

  Obviously the sum (22) is the same as in Eq.\ (18)! The correspondence
  suggests to interpret the energy spectrum (9) as representing some kind
  of surface energy associated with a quantized horizon forming a $bubble$
  of $n$ area quanta, the Planck energy $ E_{P,D}$
  playing the role of a surface tension. The presence of this energy
  guarantees the additive and equidistant quantum surface building property (11).

  Notice a qualitative difference between the droplet nucleation
  picture and the corresponding black hole horizon formation: The
  $\td$-dimensional droplets have a $(\td-1)$-dimensional boundary
  where the surface tension can be "felt", whereas the horizon
  bifurcation sphere is a closed $\td$-dimensional surface, a
  "bubble" so to speak, held together by its surface tension!

   Analogously,  the
  "orientation degeneracy field" $t=\ln g$
  appears as a "driving property", similar to an external magnetic field.

  It is worth mentioning that the above primitive Ising nucleation model
  with its purely spherical droplets
  is itself only a simplified version of the genuine Ising model which
  can describe metastable states itself \cite{le2} and which can allow
  for non-spherical excitations, too.

   The interpretation of the
  sum (22) as a grand canonical potential comes about as follows
  \cite{la1,po1,ka2}:

  One starts with a canonical partition function $Z_c$, where the same terms
  as above are summed up. However, then one has the thermodynamics of
  single $n$-droplets, but there may be $N$ of them. If one assumes these to
  be non-interacting and indistinguishable, then the grand canonical partition
  function is \bq Z_G= \sum_N \frac{Z_c^N}{N!}= e^{\ds Z_c}~, \eq which leads
  to the grand canonical potential $\psi_{\td}$ of Eq.\ (22).

   It will be
  important, however, that we interpret $Z_D$ of Eq.\ (18) as a {\it canonical}
  partition
  function, where $g=e^t$ describes the  fixed, $temperature-independent$,
  degeneracies
  of the corresponding quantum  levels, whereas
   in the droplet nucleation model $z=e^t, t= \hat{\mu}\beta$,
  is  the $temperature-dependent$ fugacity
 of a classical
  Boltzmann gas!

    Notice that $\psi_{\td}$ contains no explicit information
  about properties of the phases before and after the phase transition.
  Consider, e.g.\ a vapour $\rightarrow$ fluid phase transition. Then the
  properties of the vapour  are only very indirectly present in $\psi_{\td}$,
  namely in form of the
  surface energy $\phi$ of the droplets emerged in the vapour. The model
 here merely is supposed (for more details see the Refs.\ \cite{rev})
  to describe that  (metastable!)
 part of the
  Van der Waals isotherm
 in the $(V,p)$-plane which starts where, with decreasing volume,
 the  (theoretically!) strict equilibrium
 line of the Maxwell construction branches off to the left,
  till the (local) maximum of the Van der Waals "loop", the "spinodal" point,
 is reached.

   The series (22) converges for $t\le 0$ only.  This follows, e.g.\ from
the  Maclau\-rin-Cauchy integral criterium \cite{wh}.
 In applications to metastable systems, however,
  one is
  interested in the behaviour of $\psi_{\td}(t,x)$ for
   $t \geq 0$. This calls for
  an analytic continuation in $t$ or in the fugacity $z=e^t$ which reveals
  a branch cut of $\psi_{\td}$ from $z=1$ to $z= \infty$ \cite{fi}.

Qualitatively the following happens: For $t<0$ (i.e.\ positive
magnetic field) $\epsilon_n$ increases monotonically with $n$,
making the corresponding terms in $\psi_{\td}$ decrease
monotonically. The small droplets are favoured and no phase
transition occurs.

 If, however, $t>0$, then $\epsilon_n$ has a
maximum for \bq n^*=\left(\frac{(1-\eta)\phi)}{\hat{\mu}}\right)^{
\td} =
         \left(\frac{(1-\eta)x}{t}\right)^{ \td}~,~x=\phi\beta~, \eq with
      \bq \epsilon^* \equiv \epsilon_{n^*} =a \eta (1-\eta)^{\td -1}~,~~
      a=\frac{\phi^{\td}}{\hat{\mu}^{\td-1}}=\frac{x^{\td}}{\beta
 t^{\td -1}}~, \eq
      after which $\epsilon_n$ becomes increasingly negative with increasing
      $n$ and the series (22) explodes!

       The physical interpretation
       is the
      following: If, by an appropriate fluctuation, a "critical droplet" of
      "size" $n>n^*$ has appeared, it is energetically favored to grow. Such
 an  over-critical droplet
      -- and others of a similarly large size -- will destabilize the
 original phase and
      will send the system to the phase for which it has served as a
 nucleus!

      The energy $\epsilon^*$ may be interpreted as a measure for
      the critical barrier of the
      free energy the system has to "climb" over in order to leave the
 metastable
      state for a more stable one.

      Furthermore, the rate $\Gamma$ for the transition of the metastable
       state to the more stable one is
      proportional to $\exp(-\beta \epsilon^*)$. However, calculating the
      rate is no longer a problem of {\it equilibrium} thermodynamics. One
 has to
      deal with tools for {\it non-equilibrium} processes like the
       Fokker-Planck
      equation etc. For the droplet model this was essentially
       done by Becker and D\"oring \cite{be}. They
       assumed a stationary situation, where a steady flow of small,
      but in size increasing, droplets leave the metastable state and all
     over-critical droplets which have passed the barrier are removed from the
      system.

      Their approach was considerably improved by Langer \cite{la2}
       who related
      the transition rate $\Gamma$ to the imaginary part of $\psi_{\td}$.
      This can be seen roughly as follows:
 If one turns the sum
  $\psi_{\td}$ in Eq.\ (22) into an integral by interpreting  $n$  as a continuous
   variable:
   \bq \tilde{\psi}_{\td}=
  \int_0^{\infty}dn \, e^{\ds \beta(\hat{\mu} n -\phi\, n^{1-\eta})}~~, \eq
 then a saddle point approximation \cite{la1,ka6} for large $\beta$ gives the
 asymptotic expansion
 \bq \tilde{\psi}_{\td} \sim
  (1-\eta)^{\td/2}\sqrt{\frac{\pi\, \td}{2 \hat{\mu}\, \beta}} \left
  (\frac{\phi}{\hat{\mu}} \right)^{\td/2}
  e^{\ds -\beta a \eta(1-\eta)^{\td-1}}(i+O(1/\beta))~. \eq Here the path
 in the complex
  $n$-plane goes from $n =0$ to $ n^*$ and then parallel to the imaginary
  axis to $+i\infty$ \cite{ka6}. (Thus, only half of the associated Gaussian
 integral
  along the steepest descents contributes!)

   The crucial point is that
  the saddle point is given by $n^*$ and the associated $\epsilon^*$ of
  Eqs.\ (25) and (26), that is to
  say, by the critical droplet!

   The leading term in the saddle point
  approximation (28) is purely imaginary. Performing a Fokker-Planck type
  analysis, Langer found \cite{la2} that the transition rate $\Gamma$
  is essentially proportional to the imaginary part
 $\Im (\tilde{\psi}_{\td}),$ the other factor
  being a "dynamical" one, genuinely related to non-equilibrium properties.
  For further discussions see the reviews mentioned in
  Ref. \cite{rev}.

  An essential point for us here is the result that the imaginary part
  of $\psi_{\td}$ can be interpreted, at lest intuitively,
   in terms of equilibrium concepts although
  it is related to non-equilibrium properties which are, however, near to
  stationary situations.
 \section{Hawking temperature \\ and  Bekenstein-Hawking entropy}
 We are now ready to apply the droplet nucleation model results to
 the Schwarzschild black hole: If we denote the "critical" term in the series
 (18) by $Z^*_{D}$,  we have   \bq Z^*_{D}= e^{\ds -
 [\eta(1-\eta)^{\td-1}x^{\td}]/t^{\td-1}}~. \eq The essential point now is that
 $t=\lng$ here is no longer a temperature dependent quantity -- as in the
 droplet model -- but a fixed number. Therefore the equation of
  state for the associated
 internal
 energy, \bq U^*=-\frac{\partial \ln Z^*_D}{\partial \beta}=
 (1-\eta)^{\td-1}\left(\frac{x}{t}\right)^{\td-1} \sigma_D E_{P,D}~ = \td\,
 \epsilon^* ~, \eq
 can be used to determine  the (inverse) temperature $\beta^*$ needed
for a (potential) heat bath, if  the rest energy
 $U^*$ is given! Solving Eq.\ (30) for $x$ and using the relation (8) between
 Schwarzschild radius $R_S$ and mass $M=U^*/c^2$ we obtain
\bq \beta^*=
 \lambda \left(\frac{4\pi
R_S^*}{(\td-1)\hbar c}\right)~,~~ \lambda \equiv \frac{t\, \td\,
 \gamma}{8\pi^2}~,
\eq where $R_S^*=R_S(M=U^*/c^2)$ (Eq.\ (8)). \\ The expression in
the bracket of Eq.\ (31) is exactly the inverse Hawking
temperature in $D$-dimensional space-time \cite{pe1}, if we
 identify $U^* = Mc^2$, where $M$ is the
macroscopic rest mass of the black hole! Thus, up to a numerical
factor $\lambda$ of order 1, we obtain the Hawking
 temperature in this way. \\ For the
entropy $S^*_D=\beta^* U^* + \ln Z^*_D$ we get \begin{eqnarray}
 S^*_D/k_B
 &=& (1-\eta)^{\td}
(x^*)^{\td}/t^{\td-1}= (1-\eta)\beta^* U^* \\ & =&
 t\, n^* = \ln (g^{\ds n^*})~,~~ x^*=\beta^*\sigma_D E_{P,D}~, \nonumber
  \end{eqnarray} where
$n^*$ is the same as in Eq.\ (25).

 If we express $S^*_D$  in
terms of the $\td$-dimensional surface $A_{\td} =
\omega_{\td}\;(R_S)^{\td}$, we have \bq S^*_D/k_B = \lambda\;
\frac{A_{\td}}{4 l_{P,D}^{\td}}~ = \lambda\; \frac{c^3\,A_{\td}}{4
\hbar G_D}~~ . \eq
 So, up to the same numerical factor
$\lambda$ we  already
encountered in connection with the inverse temperature,
 we obtain the Bekenstein-Hawking
 entropy!

  The mean square fluctuations of the energy \bq
 (\Delta E)^2 = \partial^2_{\beta}(\ln Z^*_D)=-
 \frac{(1-\eta)^{\td-1}(\td-1)}{t^{\td-1}}\;
  x^{\td-2}\; (\sigma_D E_{P,D})^2 \eq
 are negative (negative specific heat!), but relatively small for large masses
 because \bq \frac{(\Delta E)^2}{U^*}=-\frac{\td-1}{\beta^*}~~, \eq
 which appears to be quite a universal relation: the r.h.s.\ of Eq.\
 (35) depends only on $\td$ and $\beta^*$!

 As the factor $\gamma$, up to now,  is a free parameter, we can possibly
choose it in such a way that
the above prefactor $\lambda$ equals one: \bq \gamma= \frac{8 \pi^2}{
t\,\td}~.
 \eq A suitable argument for such a normalization $\lambda=1$
comes from the practically classical geometrical result  $S/k_B = A_{\td}/
(4\,l^{\td}_{P,D})$ of
 Gibbons and Hawking \cite{gi1}, derived from the euclidean section of the
 Schwarzschild
 solution. In view of the surprising approximate equalities of the classical
 and the quantum theoretical values for $S$, one may use the classical result
 for normalizing the quantum one.

  There is  a corresponding analogue in QED,
  where the physical value
 of the electric charge $e$ in the quantum theory (to all orders) is
 normalized via
  the universal classical total
 Thomson  cross section $\sigma_{tot} =(8/3)\,\pi r_0^2, \; r_0= e^2/(mc^2),$
  for Compton scattering in the limit of vanishing photon energy
   \cite{qed}.

   There is a very similar universal low energy absorption cross
   section for spherically symmetric black holes \cite{das},
   namely the classical area of the horizon, which serves the same
   purpose as the Gibbons-Hawking normalization mentioned above.

  One has to be
 careful here, however, because only $Z^*_D$ has the same simple exponential
 form as one finds in the Gibbons-Hawking approach in lowest order.
 Fluctuations lead to  prefactors with powers of $x$ in front of $Z^*_D$
 as can already be seen if we take
 the imaginary part $ Z_{i,D}$ of the saddle point approximation (28) as
a slightly more sophisticated "pseudo" partition function, or, if we take
the imaginary
part of the purely imaginary partition function for the (euclidean)
Schwarzschild black hole if one includes ``quadratic'' quantum fluctuations
around the classical solution \cite{gi2,pe2,ka2}.
 Such additional powers of $x$
lead to corrections to $\beta$ and to
 logarithmic corrections of the entropy (33)
\cite{ka1,ka2}.

 As to possible  connections between  statistically defined
entropies and the ones obtained geometrically   and as to possible
quantum
 corrections
see the recent review by Frolov und Fursaev \cite{fr1}.

 The
corresponding normalization problem in loop quantum gravity has
been discussed in Refs.\ \cite{as1}.

 The normalization (36) leads
to the expression \bq a_{D-2}=4 \, (\lng)\,l^{\td}_{P,D} \eq for
the area quantum of Eq.\ (11). \\ I
 said "pseudo"
 partition function because it implies negative
mean square fluc\-tu\-ati\-ons, see Eq.\ (34) and Refs.\
\cite{ka1,ka2}, a property which
 appears to be associated
 with the
metastability of the system. More theoretical work seems to be
 necessary in order to
understand these partially surprising features better beyond the
realm of strict equilibrium thermodynamics \cite{le1}.

Altogether we see
 the following picture
emerging in the framework of the approach advocated here:

 The
formation (nucleation) of a Schwarzschild black hole means quantum
mechanically
 that the horizon
consists of a (huge) sum of Planck-sized area elements each of
which carries two Ising ($=Z(2)$)-like degrees of freedom related
to the two possible orientations. These additional degrees of
freedom lead to the Bekenstein-Hawking entropy and to the
associated Hawking temperature. The general picture is not new -
see the Refs.\ in ch.\ 3 above -, new is the suggestion that the
Ising degrees of freedom are associated with the geometrical
property "orientation" and that one should take the Ising model
seriously here because of the geometrical $Z(2)$ freedom of
orientation.

This suggestion should, of course, apply to other black holes as
well. Take, for instance, the Reissner-Nordstr{\o}m one, where the
value of the entropy for the extreme limit ($Q=M$) is
controversial: There are arguments for a vanishing entropy
\cite{haw1,te,gi3,haw2,li,br} and against \cite{za,ki}, especially
in string theory which relies heavily on the extremal property
\cite{str,ma,ho}. At present I have only a very tentative comment
because the quantum mechanics of Reissner-Nordstr{\o}m black holes
is still in a preliminary state \cite{ti,lou,br,mak,br1,bre,ki}.

Consider the submanifold characterized by $r_-=M-\sqrt{M^2-Q^2}
\le r \le r_+=M+\sqrt{M^2-Q^2}$ in the non-extreme case. The
boundaries at $r=r_-,\, r_+$ are oriented in such a way that the
normal vectors point outward relative to the manifold at both
boundaries, i.e.\ they point in opposite directions (see the
Appendix). If we now
 take the limit $r_-
\rightarrow  r_+ ~(Q \rightarrow M)$ then we have the
 singular coincidence of two spheres
with opposite orientations. As the temperature $T= (r_+^2-Q^2)/(4\pi r_+^3)$
 vanishes in the limit $Q \rightarrow M$, one may be tempted to expect
 $S\rightarrow 0$ if there is "spontaneous orientation" at $T=0$!

 "Orientation" is classically a global geometrical
  property, which means that it is
 related to the "topology" of the geometry in question. The importance of
 this aspect for the entropy of black holes has been stressed in a number
 of recent papers \cite{haw1,te,gi3,haw2,li,va1,va2,ep,haw3}.

It is remarkable that the {\em quantum} statistics of Schwarzschild black
holes is formally  the  same as that of the
{\em classical} primitive Ising droplet nucleation model
 in $D-2$ space dimensions,
 represented as a classical grand canonical ensemble.
 This shows how the
 holographic principle \cite{tho1,su} is implemented here.
  It  further indicates how the
 quantum "background" is hidden behind a classically appearing facade, very
 probably formed
 by the thermal physics of the horizon, which is being built up during the
  nucleation of the black hole.

   The "blurring" of the quantum properties
  is also indicated by the fact that the
  traces of the
  Bose statistics of the quanta (1) are rather hidden,
  contrary to, e.g., the canonical
  partition function of the simple harmonic oscillator:

   If we look at the
   exact expression for $Z_4$ in closed form, derived in Ref.\ [1],
  \begin{eqnarray} Z_4(t, x)&=&
\int_0^{\infty}d\tau
  \hat{K}(\tau, x)\frac{1}{1-e^{\ds (t-\tau)}}~,~\\ & &
\hat{K}(\tau,x) =
   \frac{x}{2\sqrt{\pi \tau^3}}e^{\ds -x^2/(4\tau)}~, \nonumber \\
      \Re [Z_4(t,x)]&= &\mbox{p.v.} \int_0^{\infty}d\tau\,
      \hat{K}(\tau,x)
   \frac{1}{1-e^{\ds t-\tau}}~,~~
\Im [Z_4(t,x)] = \pi
  \hat{K}(t,x)~,
  \nonumber \end{eqnarray} only the factor $1/(e^{t-\tau}-1)$
   in the principal
 value integral for the {\it real} part of $Z_4$ indicates Bose statistics,
 whereas the - for our discussion above crucial - imaginary part does not show
  such
traces (see also the next chapter). This property may shed new
light on  the information loss problem \cite{ha}.

  The role of the external magnetic field (or a corresponding chemical
  potential) in the droplet case  finds its correspondence in the degeneracy
  factor $g$ which intuitively represents the "driving property" leading
  to the formation of the black hole by nucleation.

   What is really
  new, compared to the droplet model,
   is that the free energy barrier $\epsilon^*$ determines its own
  temperature, namely $T_H$, and the associated entropy (32),
   as a function of the total internal energy $U^* =Mc^2$.
   This is a genuinely quantum mechanical effect,
  because $t= \ln g$ is a fixed number, not a temperature-dependent quantity
  as in the droplet nucleation case. Here lies the real difference.
  This temperature, after nucleation, then becomes that of the black hole
  itself. Any thermal radiation emitted from the horizon carries the imprint
  of this temperature.

   There is an interesting
   similarity to the Bose-Einstein
  condensation, where the critical temperature is determined as a function of
  the particle number density at vanishing chemical potential \cite{fe}.

    In the above physical interpretation of the
 nucleation process concerning the black hole I followed the droplet nucleation
 picture and have assumed that the  black hole is the $result$ of the decay
 of metastable states to a more stable one (the black hole) which
 then Hawking radiates with the corresponding temperature.

  Instead one
  might think of another
interpretation, namely that the black hole itself is the metastable
 state which
 slowly decays, due to its interaction with the heat bath consisting of
 Hawking radiation. However, here the associated Stefan-Boltzmann evaporation
 time scale $\tau$ is -- for $D=4$ -- proportional
 to $R_S^3$ (see, e.g., Ref.\ \cite{wa}), not proportional to
 $R_S$ as assumed in Eq.\ (6). Generalizing the Stefan-Boltzmann law to $D$
 space-time dimensions yields $\tau \propto R_S^{D-1}$ accordingly.
Thus, the "evaporation" interpretation is not appropriate here.

 The nucleation of black holes has been discussed
  quite early by Gross, Perry and Yaffe in terms of the euclidean Schwarzschild
  instanton \cite{pe2,ka2}.

   One can arrive at similar results for the
 Hawking
  temperature and the Be\-kenstein-Hawking entropy as above if one performs a
  microcanonical counting of states (see Ref.\ \cite{sch} and the Eqs.\
  (15)-(17) above), however, then one
   loses the very
  inspiring connection to the droplet nucleation picture.
  \section{Effective quantum \\ and  classical Hamiltonians} The partition
  function $Z_D$ of Eq.\ (18) may be rewritten as \bq Z_D = \mbox{tr}(e^{\ds -
  \beta \hat{H}})~,~\hat{H}= -\mu a^+a + \epsilon (a^+a)^{1-\eta},~
   \mu=t/\beta,~
  \epsilon = \alpha_D E_{P,D}~, \eq where $a$ and $a^+$ are the annihilation
  and creation operators of the harmonic oscillator. If $u_n$ is an
  eigenfunction of the harmonic oscillator, then $a^+a\,u_n = n\, u_n$ and the
  assertion (38) follows immediately. As the trace is independent of the basis
  one uses for its calculation, one might also use another one, e.g.\ the
  coherent states $|z \rangle$, which are eigenstates of $a$ with
   complex eigenvalues
  $z$. Doing so \cite{lo} for $\td =2$ leads to the same exact result as in
  Ref.\ [1].

   It is interesting to look \cite{mu3} briefly
   at the corresponding
 classical
       effective system: Let us define the classical
       quantity \bq \tilde{N}=\frac{1}{\hbar \omega_0}
       \left(\frac{1}{2m}p^2+\frac{m}{2}\omega_0^2 q^2\right)~. \eq  After an
       appropriate rescaling of $q$ and $p$ we have the effective Hamiltonian
  \bq \tilde{H} = -\mu \tilde{N}+ \epsilon
  \tilde{N}^{1-\eta}~,~~\tilde{N}=\frac{1}{2}(p^2+q^2)~,\eq leading to
  \begin{eqnarray}\dot{p}& =& - \frac{\partial \tilde{H}}{\partial q}=
  -\tilde{\omega}\,q~,~~\tilde{\omega}(\tilde{N})=-\mu+(1-\eta)\epsilon
 \tilde{N}^{\ds
  -\eta}~~, \\ \dot{q}&=& \frac{\partial \tilde{H}}{\partial p}=
  \tilde{\omega}\, p~. \end{eqnarray} It follows that $\tilde{N}$ is a constant
  of motion $ \tilde{N}_0$ for the associated fictitious point particle
   and that this
  particle moves on a circle with radius $\sqrt{2\tilde{N}_0}$ in phase space
 with frequency $\tilde{\omega}$
  which is a function of $\tilde{N}_0$ (this is a new feature
  compared to the usual harmonic oscillator).

   The critical value
  $\tilde{\omega}= 0$ results if $\tilde{N}=\tilde{N}_c=((1-\eta)
\epsilon/\mu)^{\td}$
  which is just the same as $n^*$ from Eq.\ (25) above and for which
 $\tilde{H}$ has its
  maximum. For $\tilde{N}<\tilde{N}_c$ the frequency $\tilde{\omega}$ is
 positive and
  for $\tilde{N}>\tilde{N}_c$ it is negative.

   As \bq \dot{q} =
  ((1-\eta)\epsilon \tilde{N}^{-\eta} -\mu)p~, \eq it is in general not at
  all trivial to calculate the Lagrange function $L(q,\dot{q})$, because one
  has to solve an algebraic equation if one wants $p(\dot{q})$ from Eq.\ (44).
  Already for $\td =2$ this equation is of order 4.
   If $\mu =0$ we get in this case
  \bq L(q,\dot{q})=-q(\frac{1}{2}\epsilon^2-\dot{q}^2)^{1/2}~, \eq
  which may be interpreted as a simple example of a Born-Infeld type
  La\-grange\-an \cite{bor}.

   Finally I mention how the classical partition
 function
 $Z_{cl}$ associated with the Hamiltonian (41) looks like: \bq
 Z_{cl}=\int_{-\infty}^{\infty}\int_{-\infty}^{\infty}
\frac{dpdq}{2\pi \hbar} \exp\{-\beta
[-\frac{\mu}{2
 \hbar}(p^2+q^2)+\frac{\epsilon}{\sqrt{2\hbar}}(p^2+q^2)^{1/2}]\} ~. \eq
 Introducing polar coordinates in the $(q,p)$-plane and making appropriate
 substitutions yields \bq Z_{cl}= 2\int_0^{\infty}du u e^{\ds t u^2-xu}~,~
 t=\beta \mu,~x=\epsilon \beta~~. \eq The integral exists for $t<0$ and
 gives \cite{GR1} \bq Z_{cl}=-\frac{e^{\ds
 -x^2/(8t)}}{t} D_{-2}(\frac{x}{\sqrt{-2t}})~,\eq where $D_p(z)$ is the
 parabolic
 cylinder function of order $p$. Continuing now from  negative to
 positive $t$ finally yields \bq
 Z_{cl}=-\frac{1}{t}\Phi(1,1/2;-\frac{x^2}{4t})+i\frac{\sqrt{\pi}x}{2t^{3/2}}
 e^{\ds -x^2/(4t)}~, \eq where $\Phi(a,c;z)$ is the confluent hypergeometric
 function \cite{er} with $\Phi(a,c;z=0)=1$.

  What is remarkable
 here is that the {\em imaginary} part
 of the classical partition function $Z_{cl}$ is the same as that of the
 quantum theoretical one (38). The real parts are different (put $x=0$),
 however, reflecting the difference between classical and quantum mechanics.
This shows again that in the context of the quantum statisics of
black holes {\em dissipative} properties of the system are
important \cite{ka3,tho4}!
  The "invariance" of the imaginary part under quantization reminds one of
  the Rutherford scattering cross section for charged particles which is
  the same in mechanics and quantum mechanics,
   due to the long range of the Coulomb forces.
\section{Conclusions} The considerations above reveal a coherent picture
of the quantum theory of isolated Schwarzschild black holes and
the associated quantum statistics:

 The first basic feature is
that the quantum horizon area spectrum is additive and equidistant
-- like the spectrum of the harmonic oscillator -- and this
universally so in any space-time dimension $D \ge 4$, the basic
quantum being of the order of $l^{D-2}_{P,D}$.  The associated
energies $E_n$ are to be interpreted as the surface energies of a
droplet of $n$ elementary quanta.

 As the classical horizon is geometrically an orientable
sphere with two possible orientations (in any dimension!) it
appears natural
 to assign this $Z(2)$ degree of freedom
to each single area quantum separately. This leads to a degeneracy
$d_n=2^n$ of the $n$-th level.

The associated quantum mechanical states consist of vectors of one
of the Hilbert spaces for (positive discrete series) irreducible
unitary representations of the group $SO^{\uparrow}(1,2)$ or its
covering groups \cite{bo1}, the $n$-quanta (area) states being
tensorially multiplied by a wave function for $n$ Ising variables
representing a  configuration  of the two possible orientations of
each of the n area quanta.

 The resulting canonical quantum statistics is
formally the same as that of the classical grand canonical
statistics of the Ising nucleation model in first-order phase
transitions for metastable states: When the droplet of $n$ area
quanta reaches a critical size the Schwarzschild system turns into
a new phase, that of a black hole.

 The imaginary part of the
complex partition functions yields Hawking's relation between
temperature and macroscopic mass of the black hole and
Bekenstein's relation between its entropy and the macroscopic size
of its horizon. The associated normalization problem for the
Hawking temperature and the Bekenstein-Hawking entropy may be
solved by refering to the essentially classical euclidean
partition function approach of Gibbons and Hawking.
\section*{Acknowledgments}
I thank M.\ Bojowald, N.\ D\"uchting and T.\ Strobl for many
discussions and my wife Dorothea for her unbelievable patience and
support.
\section*{Appendix} \section*{Geometrical orientation of manifolds}
This appendix is intended to point out some mathematical literature
dealing with the notion of "orientation" of a manifold and to indicate some
of its properties. A neat introduction into the subject
is given in chapters 6 and 7
of Ref. \cite{ab}, see also chapter 3 of Ref. \cite{gu}, chapter 5 of
Ref. \cite{si},  chapter 4 of Ref. \cite{hi}, chapters 15 and 17
of Ref. \cite{th}, chapters 9 and 10 of Ref. \cite{cu}. As to more advanced
topics like the relationship between orientation and
 characteristic classes (Stiefel-Whitney
and Euler) see Refs. \cite{mi} and \cite{hu}. For the purpose of the present
paper it is important that $n$-dimensional spheres, $n\geq 1$, are orientable
\cite{sph}. As the Kruskal-Schwarzschild and the extended
Reissner-Nordstr{\o}m manifolds may be covered by one coordinate patch
\cite{el} on which $\mbox{det}(g_{\mu \nu})$ does not vanish anywhere, they are
orientable, too.

 \end{document}